\documentstyle[12pt,epsfig]{article}
\begin{document}
\topskip 2cm
\begin{titlepage}
\rightline{ \large{ \bf 15 February 2001} }
\rightline{ \large{ \bf hep-ph/0103014} }
\begin{center}
{\large\bf The Threshold Expansion of  the 2-loop Sunrise Selfmass
         Master Amplitudes. } \\
\vspace{2.5cm}
\begin{center}
{\large {\bf
M.~Caffo$^{ab}$,
H.~Czy{\.z}\ $^{c}$}
 and
{\bf E.~Remiddi$^{ba}$ \\ } }
\end{center}

%\vspace{.5cm}
\begin{itemize}
\item[$^a$]
             {\sl INFN, Sezione di Bologna, I-40126 Bologna, Italy }
\item[$^b$]
             {\sl Dipartimento di Fisica, Universit\`a di Bologna,
             I-40126 Bologna, Italy }
\item[$^c$]
             {\sl Institute of Physics, University of Silesia,
             PL-40007 Katowice, Poland }

\end{itemize}
\end{center}

\noindent
e-mail: {\tt caffo@bo.infn.it \\
\hspace*{1.2cm} czyz@us.edu.pl \\
\hspace*{1.2cm} remiddi@bo.infn.it \\ }
\vspace{.5cm}
% \vspace{2.5cm}
% \vfil
\begin{center}
\begin{abstract}
 The threshold behavior of the master amplitudes for two loop
 sunrise self-mass graph is studied by solving the system
 of differential equations, which they satisfy.
 The expansion at the threshold
 of the master amplitudes is obtained analytically for arbitrary masses.
\end{abstract}
\end{center}
% \small{ \noindent
\scriptsize{ \noindent ------------------------------- 
}
\vfill
\end{titlepage}
\pagestyle{plain} \pagenumbering{arabic}
%%%%%%%%
\newcommand{\F}[1]{F_#1(n,m_1^2,m_2^2,m_3^2,p^2)}
\newcommand{\B}[2]{#1_#2(n,m_1^2,m_2^2,m_3^2,p^2)}
\newcommand{\D}{D(m_1^2,m_2^2,m_3^2,p^2)}
\newcommand{\A}[3]{#1_{#2,#3}(n,m_1^2,m_2^2,m_3^2,p^2)}
\newcommand{\HH}[1]{H^{(#1)}(n,m_1^2,m_2^2,m_3^2)}
\newcommand{\co}{\left(p^2+(m_1+m_2-m_3)^2\right)}
\newcommand{\GG}[1]{{\cal G}_#1(n,m_1,m_2,m_3)}
\def\Li2{\hbox{Li}_2}
\def\LLL{L(m_1^2,m_2^2,m_3^2)}
\def\a{\alpha}
\def\app{{\left(\frac{\alpha}{\pi}\right)}}
\newcommand{\Eq}[1]{Eq.(\ref{#1})}
 \newcommand{\labbel}[1]{\label{#1}}
\newcommand{\cita}[1]{\cite{#1}}
\newcommand{\dnk}[1]{ \frac{d^nk_{#1}}{(2\pi)^{n-2}} }
\newcommand{\e}{{\mathrm{e}}}
\newcommand{\verso}[1]{ {\; \buildrel {n \to #1} \over{\longrightarrow}}\; }
%%%%%%%%%%%%%%%%%%%%%%%%%%%%%%%%%%%%%%%%%%%%%%%%%%%%%%%%%%%%%%%%%%%%%%%%

\newcommand{\G}[1]{ {\cal G}_{#1}(n,m_1,m_2,m_3)}
\newcommand{\GP}[1]{ {\cal G}^{#1}_0(m_1,m_2,m_3)}

\newcommand{\GPI}[2]{ {\cal G}^{#1}_{#2}(m_1,m_2,m_3)}
\newcommand{\aaa}{ a(m_1,m_2,m_3)}
%%%%%%%%%%%%%%%%%%%%%%%%%%%%%%%%%%%%%%%%%%%%%%%%%%%%%%%%%%%%%%%%%%%%%%%%

\section{Introduction.}
The sunrise graph (also known as sunset or London transport diagram)
appears naturally, as a consequence of tensorial reduction,
in several higher order calculations in gauge theories. Due
to the presence of heavy quarks, vector bosons and Higgs particles
all the internal lines may carry different masses, so that
sunrise amplitudes depend in general on three different internal masses
\( m_i,\ i=1,2,3, \) besides the external scalar variable \( p^2 \), if
\( p_\mu \) is the external momentum (in \(n\)-dimensional Euclidean space).

For a proper understanding of their analytical behaviour, as well as for a
check of the numerical calculations, it is convenient to know the amplitudes
off-shell, and also around some particular values of \( p^2 \),
such as \( p^2 =0\), \( p^2 =\infty\), \(p^2=-(m_1+m_2-m_3)^2\)
(one of the pseudothresholds) and \(p^2=-(m_1+m_2+m_3)^2\) (the threshold).

This paper is devoted to the analytic evaluation of the coefficients
of the expansion of the sunrise amplitudes in \( p^2 \), at the
threshold value \( \ \  -p^2 = (m_1+m_2+m_3)^2 \equiv s_0 \ \  \).
The approach relies on the exploitation of the information contained
in the linear system of first order differential equations in \( p^2 \),
which is known to be satisfied by the sunrise amplitudes themselves
\cita{CCLR}. It is to be noted that all the above points
( \( p^2 =0,\ \infty\ ,\) threshold and pseudothresholds) correspond
to the Fuchsian points of the differential equations,
which therefore emerge as a natural tool for their discussion.

The analytic properties of Feynman diagrams at threshold and
pseudothresholds are well known, see for example \cita{IZ}.
The sunrise diagram, with different masses, has been
investigated in \cita{BBBS}, while in \cita{BDU} the values of the amplitudes
at threshold and pseudothreshold were obtained.
With the method established in \cita{S} and \cita{BS}, further, the
expansion around threshold was obtained in \cita{DS}, even if a complete
analytical result was given there only for the case of equal masses.
With the configuration space technique the expansion around threshold
was investigated also in \cita{GP}.
The expansion around one of the pseudothresholds (the others are
straightforwardly given by a cyclic permutation of the masses)
was obtained in analytical form in \cita{CCR}.

The threshold is a Fuchsian point of the system of equations
 obtained in \cita{CCLR}.
 In this case it is known \cita{I} that the master amplitudes
 can be expanded around that point as a combination of
 several terms, each equal to a leading power \(x^{\alpha_i}\)
 times a power series in \(x\), with \(x= (p^2+(m_1+m_2+m_3)^2) \),
 where the exponents \(\alpha_i\) are in general real numbers.
% \( (\alpha_i-\alpha_j) \ne {\rm integer}\) for \(i\ne j\).

When the expansions are inserted in the differential equations, the
equations become
a set of algebraic equations in the \(\alpha_i\) and in the coefficients
of the expansions;
the obtained algebraic equations can then be solved recursively,
for arbitrary value of the dimension \( n \),
once the initial conditions ( {\it i.e.} in the case considered here
the values of the sunrise amplitudes at the threshold) are given.
As those initial values are in turn functions of the masses, we find
in our approach that the initial values themselves satisfy a system of
linear differential equations in the masses.
We expand in the dimension \( n \) around \( n=4 \)
 the equations in the masses
 and solve them explicitly up to the finite part in
\( (n-4) \).
This result presented in the next section for \( n=4 \)
 and \( p^2= -(m_1+m_2+m_3)^2\) is in agreement with the
literature \cita{BDU}.

Once the initial conditions are obtained,
we look at the recursive solution of the algebraic equations for
the coefficients of the expansion in \( (p^2+s_0) \)
(our results are given in the
Section 3 ). It turns out that the
formula, expressing the second order coefficient of the \( (p^2+s_0) \)
 in the first master integral
 expansion in terms of the zeroth order values
 of the other master integrals, involves the coefficient
\( 1/(n-4) \).  Therefore the finite
part at \( (n-4) \) of the second order term in the \( (p^2+s_0) \) expansion
involve the first order terms in \( (n-4) \) of the zeroth order terms
in \( (p^2+s_0) \).
 We evaluate those first order terms in
\( (n-4) \) solving a differential equation, which they satisfy,
 obtained also in Section 3.
Fortunately, the formulae expressing the higher order coefficients of
the \( (p^2+s_0) \) expansion involve coefficients like \( 1/(n-5), 1/(n-6) \)
etc., which are finite at \( n=4 \), and can be used without further
problems for evaluating those higher order terms.
  Our results agree in the equal mass case with those obtained in \cite{DS}
(we take a numerical constant from the comparison),
 while for not equal masses the result is given
 in an analytical form for the first time.

%%%%%%%%%%%%%%%%%%%%%%%%%%%%%%%%%%%%%%%%%%%%%%%%%%%%%%%%%%%%%%%%%%%%%%%%
\section{The threshold values of the master amplitudes}

\newcommand{\HP}[1]{ {\cal H}^{#1}_0(m_1,m_2,m_3)}
\newcommand{\HPI}[1]{ H^{#1}_{i}(m_1,m_2,m_3)}

It is known that the two-loop sunrise self-mass graph with arbitrary masses
\( m_1, m_2, m_3 \) has four independent master amplitudes \cita{Tarasov},
which will be referred to, as in \cita{CCLR}, by
\begin{equation}
      \F{\alpha} , \hspace{1truecm} \alpha=0,1,2,3 \ ,
\labbel{Falpha}\end{equation}
where \( n \) is the continuous number of dimensions, \( m_i, i=1,2,3 \)
the three masses and \( p_\mu \) the external \(n\)-momentum.
\( \F{0} \) is the scalar amplitude
\begin{eqnarray}
  \F{0} &=& \nonumber \\
      && {\kern-200pt} \int \dnk{1} \int \dnk{2} \;
      \frac{ 1 }
           { (k_1^2+m_1^2) (k_2^2+m_2^2) ( (p-k_1-k_2)^2+m_3^2 ) } \ ,
\labbel{F0} \end{eqnarray}
 while for \( i=1,2,3 \)
\begin{equation}
      \F{i} = - \frac{\partial}{\partial m_i^2} \F{0} \ .
\labbel{Fi}\end{equation}

 The values of the master integrals at the threshold can be obtained
 in a way  similar to the way followed for the values at the
 pseudothreshold \cita{CCR} and we will discuss
 here mostly those points of the derivation which are different from the
 pseudothreshold case. If the values at the threshold are written as

\begin{equation}
 \GG{\alpha} \equiv F_{\alpha}
 \left(n,m_1^2,m_2^2,m_3^2,p^2=-(m_1+m_2+m_3)^2\right)
 \ \ , \ \alpha =0,1,2,3
 \labbel{valthr} \end{equation}

 we find, expanding around \( n=4\),

 \begin{eqnarray}
 \G{\alpha} &=& C^2(n)\Bigl[
  \frac{1}{(n-4)^2}\GPI{(-2)}{\alpha} + \frac{1}{(n-4)}\GPI{(-1)}{\alpha}
   \nonumber \\
   &&+ \GPI{(0)}{\alpha} + O(n-4)\Bigr] \ \ \  \alpha=0,1,2,3 \ .
 \labbel{6} \end{eqnarray}

 The coefficient \(C(n)\) is a function of \(n\) only
and at \(n=4\) has the following expansion
\begin{eqnarray}
C(n) &=& \left(2 \sqrt{\pi} \right)^{(4-n)} \Gamma\left(3-\frac{n}{2}\right)
\nonumber\\
     &=& 1 - (n-4) \left[ \log(2 \sqrt{\pi}) - \frac{1}{2} \gamma_E
          \right] \nonumber\\
        &+&(n-4)^2 \frac{1}{2} \left[ \log^2(2 \sqrt{\pi})
    +\frac{1}{4} \left(\frac{\pi^2}{6} + \gamma_E^2 \right)
    -\gamma_E \log(2 \sqrt{\pi}) \right] \nonumber\\
    &+& {\cal O}\left( (n-4)^3 \right) \ ,
\labbel{cn} \end{eqnarray}
where \( \gamma_E \) is the Euler-Mascheroni constant.

From \cita{CCLR}, where the singular parts in \((n-4)\)
  of \( \F{\alpha} \) are given for
arbitrary values of \( p^2\), we have
\begin{eqnarray}
   &&{\kern-20pt}\GP{(-2)} = -\frac{1}{8}(m_1^2+m_2^2+m_3^2)  \ , \nonumber \\
  &&{\kern-20pt}\GP{(-1)} = -\frac{1}{32}(m_1+m_2+m_3)^2
           + \frac{3}{16}(m_1^2+m_2^2+m_3^2) \nonumber \\
    &&{\kern+90pt} -\frac{1}{8}\left[
      m_1^2\log(m_1^2)+m_2^2\log(m_2^2)+m_3^2\log(m_3^2) \right] \ ,
 \nonumber \\
 &&{\kern-20pt}\GPI{(-2)}{i} = \frac{1}{8} \ \ ,
 \ \ \GPI{(-1)}{i} = -\frac{1}{16}
  +\frac{1}{8}\log(m_i^2) \ \ , \ \ i=1,2,3.
 \labbel{7} \end{eqnarray}

To obtain the other coefficients of \Eq{6}, as in \cita{CCR},
our starting point is
the  system of four linear differential equations in $ p^2 $,
satisfied by the four master amplitudes
$ F_{\alpha}(n,m_1^2,m_2^2,m_3^2,p^2)$, $\alpha =0,1,2,3$,
which are not rewritten here for short.
By imposing the condition \Eq{valthr} to $ p^2 $ the system becomes
equivalent to a system of three linear differential equations in $m_3$,
which in turn can be written as a single third order differential equation
in $m_3$ for the function \( \GPI{(0)}{0} \).
We do not repeat here the details of the derivation, which is
very similar to the derivation leading to the analogous
Eq.(17) of \cita{CCR} (with the important difference of the
substitution $ m_3 \rightarrow -m_3\ $).
Its solution can also be obtained again in an analogous way, for
positive values of \(m_1\),\(m_2\) and \(m_3\); it contains three
unknown constants of integration (indeed functions of the masses $m_1$
and $m_2$) \({{\cal C}_i(m_1,m_2)}, i=1,2,3\),
to be fixed by the initial conditions.

 The initial conditions have to be imposed however in a way different with
 respect to the pseudothreshold.
 The point \(p^2=0\), which is known \cita{CCLR}
 and was used to impose initial conditions at the pseudothreshold \cite{CCR},
 corresponds in the case of the threshold to the value \(m_3=-(m_1+m_2) \),
 which is outside the validity of the present solution for \( \GPI{(0)}{0} \),
 where positive values of the masses are assumed.
 However we know already the value of the function at the pseudothreshold
 \cita{BDU,CCR} and for \(m_3=0\) the threshold and pseudothreshold are
 identical. That allows us to choose that point for the initial conditions.
 Imposing that, at \(m_3=0 ,\ \ \) \( G_0^{(0)}(m_1,m_2,m_3) \),
 given by the Eq.(25) of \cita{CCR}, and its first derivative with respect
 to \(m_3\) are equal to the function
 \( \GP{(0)} \) and to its first \(m_3\) derivative respectively,
 one gets two relations between the three functions
 \({\cal C}_i(m_1,m_2)\), \(i=1,2,3\).
 The second derivative at that point is infinite and does not give
 an additional relation. After eliminating the two coefficients
 \({\cal C}_i(m_1,m_2)\), \(i=2,3\) one gets \( \GP{(0)}  \)
 as a function of \({\cal C}_1(m_1,m_2)\) only. The remaining constant can
 be fixed by the relations obtained by the permutation of the masses as we
 know that the \( \GP{(0)} \) is symmetric under that exchange.
 Once all the constants are fixed the solution reads:

\begin{eqnarray}
 &&\GP{(0)} = \nonumber \\
   &&-\frac{1}{8(m_1+m_2+m_3)^2}
   \Biggl[ m_1^3(m_1+2m_2){\cal L}_t(m_1,m_2,m_3)
         +m_2^3(2m_1+m_2){\cal L}_t(m_2,m_1,m_3)\nonumber \\
  &&{\kern100pt}-2\pi\left(m_1 m_2 +m_1 m_3 + m_2 m_3\right)
   \sqrt{m_1 m_2 m_3 (m_1+m_2+m_3)}
  \Biggr] \nonumber \\
   &&+\frac{1}{4(m_1+m_2+m_3)}
   \Biggl[ (m_1+m_2)\left(m_1^2{\cal L}_t(m_1,m_2,m_3)
         +m_2^2{\cal L}_t(m_2,m_1,m_3)\right) \nonumber \\
   &&{\kern100pt}
  +\frac{1}{2}(m_1^2+m_2^2+m_1m_2)\left(m_1\log\left(\frac{m_3}{m_1}\right)
     +m_2\log\left(\frac{m_3}{m_2}\right) \right)        \Biggr]  \nonumber \\
 &&+\frac{1}{8}(m_3^2-m_1^2-m_2^2)\left({\cal L}_t(m_1,m_2,m_3)
         + {\cal L}_t(m_2,m_1,m_3)\right) \nonumber \\
  &&-\frac{1}{32}\Bigl[ m_1^2\log^2(m_1^2)+ m_2^2\log^2(m_2^2)
  + m_3^2\log^2(m_3^2)
 +(m_1^2+m_2^2-m_3^2)\log(m_1^2)\log(m_2^2) \nonumber \\
 &&+(m_1^2-m_2^2+m_3^2)\log(m_1^2)\log(m_3^2)
 +(-m_1^2+m_2^2+m_3^2)\log(m_2^2)\log(m_3^2) \nonumber \\
 &&-m_1(7m_1-2m_3)\log(m_1^2)
 -m_2(7m_2-2m_3)\log(m_2^2) \nonumber \\
 &&+(2m_1^2 + 2m_1m_2 + 2m_2^2 -5m_3^2)\log(m_3^2)\Bigr]
  -\frac{\pi}{4} \sqrt{m_1 m_2 m_3 (m_1+m_2+m_3)}\nonumber \\
  &&
 -\frac{11}{128}(m_1^2+m_2^2+m_3^2) +\frac{13}{64}(m_1m_2+m_1m_3+m_2m_3)
\labbel{V17} \end{eqnarray}
where
\begin{eqnarray}
 {\cal L}_t(m_1,m_2,m_3) &=& -\hbox{Li}_2\left(-\frac{m_3}{m_2}\right)
                          -\hbox{Li}_2\left(-\frac{m_1}{m_2}\right)
             + \log\left(\frac{m_3}{m_1+m_2}\right)
                \log\left(\frac{m_1}{m_2}\right)\nonumber \\
 && {\kern-100pt}- \log\left(\frac{m_3}{m_2}\right)
      \log\left(\frac{m_2+m_3}{m_2}\right) -\frac{5}{6}\pi^2
  +2\pi
 \arctan\left(\sqrt{\frac{m_2(m_1+m_2+m_3)}{m_1m_3}} \right) \ .
\labbel{V18} \end{eqnarray}
The above result is in agreement with \cite{BDU}. To check it one has
to expand in \( n \) around \(n=4\) all the factors which
depend on \( n \), such as our \( C(n) \) ( \Eq{cn})
and the \( n \)-dependent functions of \cite{BDU}.
\par
 The functions \( \G{i} \ \ \  i=1,2,3\) can be easily found
 in an analogous way as for the pseudothreshold \cite{CCR}
 and the solutions read
\begin{eqnarray}
 &&\G{1} = \nonumber \\
 &&\frac{1}{8m_1 m_2} \Biggl\{
  -\left(m_1+3m_2\right) \frac{\partial}{\partial m_1} \G{0}
  +\left(m_2-m_3\right) \frac{\partial}{\partial m_3} \G{0}\nonumber \\
  &&+\frac{1}{m_1+m_2+m_3}\Biggl[
 \left(\left(n-3\right)\left(2m_1+m_2+2m_3\right)-m_2\right)\G{0}\nonumber \\
 && + \frac{(n-2)^2}{2(n-3)}\ \ \ \Biggl(\frac{1}{m_1}T(n,m_1^2)T(n,m_2^2)
  +\frac{m_2}{m_1 m_3}T(n,m_1^2)T(n,m_3^2)\nonumber \\
  &&{\kern100pt}+\frac{1}{m_3}T(n,m_2^2)T(n,m_3^2)\Biggr)\ \ \ \Biggr]\Biggr\}
\labbel{V19} \end{eqnarray}
\begin{eqnarray}
 &&\G{2} = \nonumber \\
 &&\frac{1}{8 m_2^2} \Biggl\{
   \left(3m_1+m_2\right) \frac{\partial}{\partial m_1} \G{0}
  +\left(m_2+3m_3\right) \frac{\partial}{\partial m_3} \G{0}\nonumber \\
  &&+\frac{1}{m_1+m_2+m_3}\Biggl[
 -\left(\left(n-3\right)\left(6m_1+7m_2+6m_3\right)+m_2\right)\G{0}\nonumber \\
 && + \frac{(n-2)^2}{2(n-3)}\ \ \ \Biggl(\frac{1}{m_1}T(n,m_1^2)T(n,m_2^2)
  +\frac{m_2}{m_1 m_3}T(n,m_1^2)T(n,m_3^2)\nonumber \\
  &&{\kern100pt}+\frac{1}{m_3}T(n,m_2^2)T(n,m_3^2)\Biggr)\ \ \ \Biggr]\Biggr\}
\labbel{V20} \end{eqnarray}
\begin{eqnarray}
 &&\G{3} = \nonumber \\
 &&\frac{1}{8m_2 m_3} \Biggl\{
  -\left(m_1-m_2\right) \frac{\partial}{\partial m_1} \G{0}
  -\left(3m_2+m_3\right) \frac{\partial}{\partial m_3} \G{0}\nonumber \\
  &&+\frac{1}{m_1+m_2+m_3}\Biggl[
 \left(\left(n-3\right)\left(2m_1+m_2+2m_3\right)-m_2\right)\G{0}\nonumber \\
 && + \frac{(n-2)^2}{2(n-3)}\ \ \ \Biggl(\frac{1}{m_1}T(n,m_1^2)T(n,m_2^2)
  +\frac{m_2}{m_1 m_3}T(n,m_1^2)T(n,m_3^2)\nonumber \\
  &&{\kern100pt}+\frac{1}{m_3}T(n,m_2^2)T(n,m_3^2)\Biggr)\ \ \ \Biggr]\Biggr\}
 \labbel{V21} \end{eqnarray}
 where \( T(n,m^2) \) is defined by
\begin{equation}
  T(n,m^2) = \int\dnk{ } \frac{1}{k^2+m^2} = \frac{m^{n-2}}{(n-2)(n-4)}C(n)\ .
\labbel{Tnm} \end{equation}

 After expanding around \(n=4\) one gets, besides of
 \Eq{7},
\begin{eqnarray}
 &&\GPI{(0)}{3}  = \frac{1}{8(m_1+m_2+m_3)^2}
   \Biggl[2\pi \sqrt{m_1 m_2 m_3(m_1+m_2+m_3)}\nonumber \\
   && {\kern+180pt}+m_1^2{\cal L}_t(m_1,m_2,m_3)
         +m_2^2{\cal L}_t(m_2,m_1,m_3)
    \Biggr]\nonumber \\
  &&+\frac{1}{8(m_1+m_2+m_3)}\Biggl[
    m_1\log\left(\frac{m_1}{m_3}\right)+m_2\log\left(\frac{m_2}{m_3}\right)
    \Biggr]\nonumber \\
  &&-\frac{1}{8}\Biggl[ {\cal L}_t(m_1,m_2,m_3)+{\cal L}_t(m_2,m_1,m_3)
 -\log^2\left(m_3\right)\nonumber \\
  &&-\log\left(m_3\right)\log\left(m_1\right)
 -\log\left(m_3\right)\log\left(m_2\right)
 +\log\left(m_1\right)\log\left(m_2\right)
 +\log\left(m_3\right)+\frac{1}{4}\Biggr]
 \ . \labbel{V22} \end{eqnarray}

 The other two functions, \( \GPI{(0)}{1}\) and \( \GPI{(0)}{2}\),
 can be easily obtained by a permutation of the masses
 and we do not report them here. Again the results are in agreement with
\cite{BDU}.
%%%%%%%%%%%%%%%%%%%%%%%%%%%%%%%%%%%%%%%%%%%%%%%%%%%%%%%%%%%%%%%%%%%%%%%%
 \section {The expansion at the threshold.}

\newcommand{\HC}[1]{ {\cal H}^{#1}(n,m_1,m_2,m_3)}
\newcommand{\HS}[1]{ {\cal H}_s^{#1}(n,m_1,m_2,m_3)}

 The expansions at the threshold of the master amplitudes can be found
 with the help of the system of equations
 obtained in \cite{CCLR} (Eqs.(5,7)). As the threshold is a singular point
 the expansions consist of the regular parts plus the singular ones
 with fractional exponents. By inserting the expansion into
 the system of equations one finds that the
 fractional powers are fixed and
 the expansion reads

\begin{eqnarray}
 \F{\alpha} &&= \sum_{i=0}^{\infty}
 \HC{(\alpha,i)} x^i \nonumber \\
 &&{\kern-100pt}+ x^{n-j(\alpha)}\HS{(\alpha,0)}
  \left(1+\sum_{i=1}^{\infty}\HS{(\alpha,i)}x^i\right)
  \ , \labbel{EX1} \end{eqnarray}

\noindent
 where \(n\) is the continuous dimension, \(\alpha =0,1,2,3\);
 \(j(0)=2, \ \ j(1)=j(2)=j(3)=3 \) and
\begin{equation}
 x = p^2+(m_1+m_2+m_3)^2
  \  \labbel{EX1A} \end{equation}
\noindent
 is the expansion parameter.
Of course in this new notation
\begin{equation}
 \HC{(\alpha,0)} \equiv \GG{\alpha}
  \  \labbel{equiv} \end{equation}
\noindent
is already presented in \Eq{6}.

 From \Eq{Fi} it
 follows that the \(\HC{(\alpha,i)} \) and
 \(\HS{(\alpha,i)} \) are related, so that having
 \(\HC{(0,i)} \) and \(\HS{(0,i)} \) one can calculate the others
 for \(\alpha =1,2,3\).

 From the discussed system of equations
 follows also that the following differential equations
 hold for \(\HS{(0,0)} \)

\begin{eqnarray}
  &&\frac{\partial \HS{(0,0)}}{\partial m_1} =
  \Biggl[ \frac{n-3}{2 m_1} + \frac{5-3 n}{2\left(m_1+m_2+m_3\right)}
   \Biggr]\HS{(0,0)}\nonumber \\
  &&\frac{\partial \HS{(0,0)}}{\partial m_2} =
  \Biggl[ \frac{n-3}{2 m_2} + \frac{5-3 n}{2\left(m_1+m_2+m_3\right)}
   \Biggr]\HS{(0,0)}\nonumber \\
  &&\frac{\partial \HS{(0,0)}}{\partial m_3} =
  \Biggl[ \frac{n-3}{2 m_3} + \frac{5-3 n}{2\left(m_1+m_2+m_3\right)}
   \Biggr]\HS{(0,0)}
  \ . \labbel{EX2} \end{eqnarray}

 That means that \(\HS{(0,0)} \) can be written as

\begin{eqnarray}
 \HS{(0,0)} = {\cal C}_{\cal H}\left(n\right) \ \ \
 \frac{\left(m_1 m_2 m_3 \right)
 ^{\frac{n-3}{2} }}
                           {\left(m_1+m_2+m_3\right)^{\frac{3n-5}{2} }}
  \ , \labbel{EX3} \end{eqnarray}

\noindent
 where \( {\cal C}_{\cal H}\left(n\right)\) is a function of \(n\) only,
 to be discussed later.
 As usual, the coefficients
 \(\HS{(0,i)}\), \( i=1,\cdots \) can be found
 by solving a system of (in this case four) linear equations.
 We report here for brevity only one of the coefficients

\begin{eqnarray}
 &&\HS{(0,1)} = \nonumber \\
 &&\frac{1}{16(m_1+m_2+m_3)}
 \Biggl[ \frac{3\left( 3n-5\right)}{m_1+m_2+m_3}
  -\left(n-3\right)
  \left(\frac{1}{m_1} +\frac{1}{m_2}+\frac{1}{m_3}\right)\Biggr]
   \ . \labbel{EX4} \end{eqnarray}

The singular part of the expansion in \Eq{EX1} can be compared
with the results of \cita{DS}, as it corresponds, for arbitrary masses
and in the language of that reference, to the (p-p) regions contribution
completely expanded at the threshold.
Our results \Eq{EX1}, \Eq{EX3} and \Eq{EX4} agree analytically with those
presented in Eq.(5), Eq.(49), Eq.(50) of \cita{DS}, provided that
the value given in Eq.(49) of \cita{DS} is taken with a minus sign,
in agreement with its equal mass limit, Eq.(51) of \cita{DS}.

 In the regular part of the expansion all the coefficients can be found
 provided we know \(\HC{(\alpha,0)}\equiv  \G{\alpha}\),
 for \(\alpha =0,1,2,3\), already given in \Eq{6}.
 This is only partly true
 as we know
\(\HC{(\alpha,0)}\) expanded around \(n=4\) up to
 the constant term only.
The next to lowest term in the  \(x\)-expansion reads

\begin{eqnarray}
 &&\HC{(0,1)} = -\frac{1}{(m_1+m_2+m_3)^2}\Bigl[ (n-3)\G{0} \nonumber \\
 &&
 +m_1^2\G{1}+m_2^2\G{2}+m_3^2\G{3} \Bigr]
   \ . \labbel{EX5} \end{eqnarray}

 The \Eq{EX5} allows us to find the expansion of \(\HC{(0,1)}\) around \(n=4\)
  up to the constant term as we know the expansion of \(\G{\alpha}\) up
 to the constant terms. It can be written as

\begin{eqnarray}
 \HC{(0,1)}= C^2(n)\left[ \frac{1}{n-4}{\cal H}_{(0,1)}^{(-1)}
  + {\cal H}_{(0,1)}^{(0)} +  O(n-4)\right]
  \ , \labbel{EX5A} \end{eqnarray}

\noindent
with

\begin{eqnarray}
 {\cal H}_{(0,1)}^{(-1)}= \frac{1}{32}
  \ , \labbel{EX5B} \end{eqnarray}

\noindent
and

\begin{eqnarray}
 &&{\cal H}_{(0,1)}^{(0)} = -\frac{1}{8(m_1+m_2+m_3)^4}
        \Biggl[
  2\pi\left(m_1^2+m_1m_2+m_2^2\right)\sqrt{m_1m_2m_3(m_1+m_2+m_3)}
 \nonumber \\
             &&{\kern+80pt}+m_1^3(m_1+2m_2){\cal L}_t(m_1,m_2,m_3)
         +m_2^3(2m_1+m_2){\cal L}_t(m_2,m_1,m_3) \Biggr] \nonumber \\
&& -\frac{1}{16(m_1+m_2+m_3)^3} \Biggl[
   -4\pi\left(m_1+m_2\right)\sqrt{m_1m_2m_3(m_1+m_2+m_3)}\nonumber \\
  &&{\kern+80pt}-4(m_1+m_2)\left(m_1^2{\cal L}_t(m_1,m_2,m_3)
                  +m_2^2{\cal L}_t(m_2,m_1,m_3)\right)
 \nonumber \\
 &&{\kern+80pt}+ \left(m_1^2+m_1m_2+m_2^2\right)
  \left(m_1\log(m_1^2)+m_2\log(m_2^2)\right) \nonumber \\
 &&{\kern+80pt}
 -\left(m_1^3+2m_1^2m_2+2m_1m_2^2+m_2^3\right)\log(m_3^2)\Biggr]\nonumber \\
 &&{\kern-17pt}-\frac{1}{32(m_1+m_2+m_3)^2} \Biggl[
  4m_1^2{\cal L}_t(m_1,m_2,m_3)+4m_2^2{\cal L}_t(m_2,m_1,m_3)
 -2(m_1^2+m_1m_2+m_2^2)\nonumber \\
 &&{\kern-17pt}-m_1(3m_1+2m_2)\log(m_1^2)-m_2(2m_1+3m_2)\log(m_2^2)
 +(3m_1^2+4m_1m_2+3m_2^2)\log(m_3^2)
   \Biggr]\nonumber \\
 &&-\frac{m_1+m_2}{16(m_1+m_2+m_3)} +\frac{1}{32}\log(m_3^2)-\frac{5}{128}
  \ . \labbel{EX5C} \end{eqnarray}

In the equal mass case (\(m_1=m_2=m_3\))
 the above result is in agreement, up to terms \( O(n-4)\),
 with \cite{DS}
 , while the analytical result
 for all masses different was not previously known in the literature.

 In order to obtain the first order term in the \(x\)-expansion for
 the other master amplitudes, we choose to
 proceed with the expansion of \(\F{0}\). It is necessary to know
 terms of the order \(x^2\) of \(\F{0}\) to find terms of the order
 \(x\) of the other master amplitudes using the relations of \Eq{Fi}.
 For the expression of \(\HC{(0,2)}\) we find

 \begin{eqnarray}
 \HC{(0,2)}&=& \frac{1}{32(n-4)} \ \ \frac{1}{(m_1+m_2+m_3)^3}\nonumber \\
 &&{\kern-120pt}\Biggl[ (2m_1+m_2+m_3)\G{1} +(m_1+2m_2+m_3)\G{2}\nonumber \\
 &&+(m_1+m_2+2m_3)\G{3}\Biggr]
  + \cdots
   \ . \labbel{EX6} \end{eqnarray}

 We do not write here explicitly other terms, which are known, containing
 up to triple pole in the \((n-4)\) expansion. In fact as one can see later
 (\Eq{EX14}) the triple and double pole terms cancel in the expansion.
 The presence of the explicitly shown term \(\sim \frac{1}{n-4}\)
 is a kind of an obstacle analogous
 to the one encountered at the pseudothreshold expansion \cite{CCR}.
 In principle it requires the knowledge of the expansion of the combination
 of the master integrals written in square brackets of \Eq{EX6} up
 to the term  \(\sim (n-4)\) included.
  We could solve that problem in an analogous
 way as in the pseudothreshold expansion \cite{CCR}, but to show how far we
 can
 get just relying on the differential equations we will proceed differently.
 Defining the above combination

   \begin{eqnarray}
  &&\G{c} = \Biggl[ (2m_1+m_2+m_3)\G{1} \nonumber \\
 &&+(m_1+2m_2+m_3)\G{2}
 +(m_1+m_2+2m_3)\G{3}\Biggr]
   \ , \labbel{EX7} \end{eqnarray}

 we calculate the \(m_3\) derivative of it

   \begin{eqnarray}
 &&{\kern-30pt}\frac{\partial \G{c}}{\partial m_3} =\nonumber \\
   &&(n-3)\G{c}
      \Bigl( \frac{3}{2m_3} + \frac{1}{2(m_1+m_2+m_3)}
            -\frac{2}{m_1+m_2+2m_3} \Bigr)\nonumber \\
   &&+\G{c}
      \Bigl( -\frac{1}{m_3} - \frac{1}{m_1+m_2+m_3}
            +\frac{2}{m_1+m_2+2m_3}\Bigr)\nonumber \\
   &&+ (n-4)\G{1}\Bigl(-1 -\frac{2m_1+m_2}{m_3}
          + \frac{3m_1+m_2}{m_1+m_2+2m_3}\Bigr) \nonumber \\
   &&+ (n-4)\G{2}\Bigl(-1 -\frac{m_1+2m_2}{m_3}
          + \frac{m_1+3m_2}{m_1+m_2+2m_3}\Bigr) \nonumber \\
   &&+\frac{(n-2)^2}{4m_2^2m_3^2}T(n,m_2^2)T(n,m_3^2)
 \Bigl(1-\frac{m_2}{2m_3} -\frac{m_1}{2(m_1+m_2+m_3)}\Bigr)\nonumber \\
   &&+\frac{(n-2)^2}{4m_1^2m_3^2}T(n,m_1^2)T(n,m_3^2)
 \Bigl(1-\frac{m_1}{2m_3} -\frac{m_2}{2(m_1+m_2+m_3)}\Bigr)\nonumber \\
   &&+\frac{(n-2)^2}{4m_1^2m_2^2}T(n,m_1^2)T(n,m_2^2)
 \Bigl(-\frac{m_1+m_2}{2m_3} +\frac{m_1+m_2}{2(m_1+m_2+m_3)}\Bigr)
    \ . \labbel{EX8} \end{eqnarray}

 When expanding around \(n=4\),  we find
 (due to the factor \((n-4)\) in front of
 \(\G{1}\) and \( \G{2}\) ) that the term of \( \G{c}\)
 proportional to \( (n-4)\), denoted as \(G_{c}^{(1)}(m_1,m_2,m_3)\),

    \begin{eqnarray}
    \G{c}\ = C^2(n)\Bigl[ \cdots + (n-4)G_{c}^{(1)}(m_1,m_2,m_3)
   + \cdots \Bigr]
  \ , \labbel{EX9} \end{eqnarray}

\noindent
 fulfills the following
 first order differential equation

   \begin{eqnarray}
 && \frac{\partial G_{c}^{(1)}(m_1,m_2,m_3)}{\partial m_3} =
  \frac{1}{2}G_{c}^{(1)}(m_1,m_2,m_3) \Bigl(\frac{1}{m_3}
          - \frac{1}{m_1+m_2+m_3} \Bigr)\nonumber \\
 &&+\frac{1}{2}\GPI{(0)}{1}\Bigl(\frac{2m_1+m_2}{m_3}+\frac{m_1}{m_1+m_2+m_3}
      \Bigr)\nonumber \\
  &&+\frac{1}{2}\GPI{(0)}{2}\Bigl(\frac{2m_2+m_1}{m_3}+\frac{m_2}{m_1+m_2+m_3}
      \Bigr)\nonumber \\
  &&+\frac{1}{2}\GPI{(0)}{3}\Bigl(
   4 + \frac{3(m_1+m_2)}{m_3}-\frac{m_1+m_2}{m_1+m_2+m_3}
      \Bigr)\nonumber \\
  &&-\frac{1}{384}\log^3(m_1^2)\Bigl(
     -2 + \frac{2m_1+m_2}{m_3}-\frac{m_1}{m_1+m_2+m_3}\Bigr) \nonumber \\
  &&-\frac{1}{384}\log^3(m_2^2)\Bigl(
     -2 + \frac{m_1+2m_2}{m_3}-\frac{m_2}{m_1+m_2+m_3}\Bigr)\nonumber \\
  &&-\frac{1}{384}\log^3(m_3^2)\Bigl(
     -4 + \frac{m_1+m_2}{m_3}+\frac{m_1+m_2}{m_1+m_2+m_3}\Bigr)\nonumber \\
  &&-\frac{1}{128}\log(m_1^2)\log(m_2^2)\Bigl(\log(m_1^2)+\log(m_2^2) \Bigr)
  \Bigl(
      \frac{m_1+m_2}{m_3}-\frac{m_1+m_2}{m_1+m_2+m_3}\Bigr)\nonumber \\
  &&-\frac{1}{128}\log(m_1^2)\log(m_3^2)\Bigl(\log(m_1^2)+\log(m_3^2) \Bigr)
  \Bigl( -2
      + \frac{m_1}{m_3}+\frac{m_2}{m_1+m_2+m_3}\Bigr)\nonumber \\
  &&-\frac{1}{128}\log(m_2^2)\log(m_3^2)\Bigl(\log(m_2^2)+\log(m_3^2) \Bigr)
  \Bigl( -2
      + \frac{m_2}{m_3}+\frac{m_1}{m_1+m_2+m_3}\Bigr)
   \ , \labbel{EX10} \end{eqnarray}

 instead of being a part of a system of three differential equations,
 as in the case of arbitrary \(n\).

 We can find a solution of that equation in a relatively simple way,
 which consists mainly in integrating by parts, so we do not report
 here details of the derivation. The solution reads

 \begin{eqnarray}
 &&{\kern-20pt} G_{c}^{(1)}(m_1,m_2,m_3) =
   \frac{(m_2-m_3)(2m_1+m_2+m_3)}{4(m_1+m_2+m_3)}
    {\cal L}_t(m_1,m_2,m_3)\nonumber \\
  &&{\kern+68pt}+ \frac{(m_1-m_3)(m_1+2m_2+m_3)}{4(m_1+m_2+m_3)}
    {\cal L}_t(m_2,m_1,m_3)\nonumber \\
   &&+\frac{1}{192}\Bigl[ \log^3(m_1^2)\left(2m_1+m_2+m_3\right)
   + \log^3(m_2^2)\left(m_1+2m_2+m_3\right) \nonumber \\
   &&{\kern+20pt}+\log^3(m_3^2)\left(m_1+m_2+2m_3\right)\Bigr]\nonumber \\
 &&+\frac{1}{64}\log(m_1^2)\log(m_2^2)\left(\log(m_1^2)+\log(m_2^2)\right)
    \left(m_1+m_2\right)\nonumber \\
 &&+\frac{1}{64}\log(m_1^2)\log(m_3^2)\left(\log(m_1^2)+\log(m_3^2)\right)
    \left(m_1+m_3\right)\nonumber \\
 &&+\frac{1}{64}\log(m_2^2)\log(m_3^2)\left(\log(m_2^2)+\log(m_3^2)\right)
    \left(m_2+m_3\right)\nonumber \\
 &&{\kern-13pt}-\frac{1}{32}\Bigl[ m_3\left(\log(m_1^2)+\log(m_2^2)\right)^2
                      +m_2\left(\log(m_1^2)+\log(m_3^2)\right)^2
                      +m_1\left(\log(m_2^2)+\log(m_3^2)\right)^2\Bigr]
     \nonumber \\
   &&+\frac{1}{8}\Bigl[\log(m_1^2)\left(-3m_1+m_2+m_3\right)
                      +\log(m_2^2)\left(m_1-3m_2+m_3\right)\nonumber \\
                      &&+\log(m_3^2)\left(m_1+m_2-3m_3\right)\Bigr]
      +\frac{11}{16}\left(m_1+m_2+m_3\right)\nonumber \\
 &&+ \sqrt{\frac{m_1m_2m_3}{m_1+m_2+m_3}}
  \Biggl\{
 \pi\Biggl[\frac{3}{4}\left(\log(m_1^2)+\log(m_2^2)+\log(m_3^2)\right)
   +\frac{1}{2}\log(m_1+m_2+m_3)\nonumber \\
 &&-\log(m_1+m_2)-\log(m_1+m_3)-\log(m_2+m_3)
     \Biggr]  +{{\cal I}_3(m_1,m_2,m_3)} -K
  \Biggr\}
    \ , \labbel{EX11} \end{eqnarray}

\noindent
 where the unknown function of \(m_1\) and \(m_2\), which remains after
 integration, is reduced to the single constant \(K\) using the symmetry
 of \(G_{c}^{(1)}(m_1,m_2,m_3)\) under the interchange of all the masses.
 The actual value of \(K\) will be fixed later.
\({\cal L}_t(m_1,m_2,m_3)\) is defined in \Eq{V18}
and the only non trivial integral left,
 \({{\cal I}_3(m_1,m_2,m_3)}\), is

 \begin{eqnarray}
 &&{\cal I}_3(m_1,m_2,m_3)=\tilde{\cal I}_3(m_1,m_2,m_3)
 +\tilde{\cal I}_3(m_1,m_1,m_2)-\tilde{\cal I}_3(m_2,m_1,m_1) \ ,
\nonumber \\
\nonumber \\
&&{\tilde{\cal I}_3(m_1,m_2,m_3)}=
 \sqrt{m_1 m_2} \int dm_3 \frac{1}{\sqrt{m_3(m_1+m_2+m_3)}}
  \left[\frac{\log\left(\frac{m_3}{m_1}\right)}{m_3+m_1}
     +\frac{\log\left(\frac{m_3}{m_2}\right)}{m_3+m_2}\right]= \nonumber \\
 && \kern-12pt i\Biggl\{\log\left(\frac{m_1+m_2}{4m_1}\right)
   \left[  \log(t-t_1)-\log(t-t_2)\right]
      +\log\left(\frac{m_1+m_2}{4m_2}\right)
     \left[\log(t+t_2)-\log(t+t_1)\right]  \nonumber \\
  &&+\log(t-t_1)\left[2\log(1-t_1)-\log(t_1)\right]
  -\log(t-t_2)\left[2\log(1-t_2)-\log(t_2)\right]\nonumber \\
  &&-\log(t+t_1)\left[2\log(1+t_1)-\log(-t_1)\right]
  +\log(t+t_2)\left[2\log(1+t_2)-\log(-t_2)\right]\nonumber \\
  &&-2\ \hbox{Li}_2\left(\frac{t-t_1}{1-t_1}\right)
    +2\ \hbox{Li}_2\left(\frac{t-t_2}{1-t_2}\right)
    +\hbox{Li}_2\left(-\frac{t-t_1}{t_1}\right)
    -\hbox{Li}_2\left(-\frac{t-t_2}{t_2}\right)\nonumber \\
  &&+2\ \hbox{Li}_2\left(\frac{t+t_1}{1+t_1}\right)
    -2\ \hbox{Li}_2\left(\frac{t+t_2}{1+t_2}\right)
    -\hbox{Li}_2\left(\frac{t+t_1}{t_1}\right)
    +\hbox{Li}_2\left(\frac{t+t_2}{t_2}\right) \Biggr\}
    \ , \labbel{EX12} \end{eqnarray}

\noindent
where \(t\) and \(t_{1,2}\) are defined as

\begin{eqnarray}
 t&=& \frac{\sqrt{m_1+m_2+m_3}-\sqrt{m_3} }{ \sqrt{m_1+m_2+m_3}+\sqrt{m_3}} \ ,
\nonumber \\
  t_{1,2} &=& \frac{m_2-m_1 \pm 2i \sqrt{m_1m_2}}{m_1+m_2}
    \ . \labbel{EX13} \end{eqnarray}

 Having \(G_{c}^{(1)}(m_1,m_2,m_3)\) we can find the \(n=4\) expansion
 of \(\HC{(0,2)}\) up to the constant term. It reads

\begin{eqnarray}
 \HC{(0,2)}= C^2(n)\left[ \frac{1}{(n-4)}{\cal H}_{(0,2)}^{(-1)}
  + {\cal H}_{(0,2)}^{(0)} +  O(n-4)\right]
    \ , \labbel{EX14} \end{eqnarray}

\noindent
where

\begin{eqnarray}
 {\cal H}_{(0,2)}^{(-1)} = \frac{\pi}{32}
  \frac{\sqrt{m_1m_2m_3(m_1+m_2+m_3)}}{(m_1+m_2+m_3)^4}
    \ , \labbel{EX15} \end{eqnarray}

\noindent
and

\begin{eqnarray}
 &&{\kern-20pt}{\cal H}_{(0,2)}^{(0)}=\nonumber \\
 &&-\frac{1}{8(m_1+m_2+m_3)^6}\Bigl[
  m_1^3(m_1+2m_2){\cal L}_t(m_1,m_2,m_3)
 +m_2^3(2m_1+m_2){\cal L}_t(m_2,m_1,m_3)
 \Bigr]\nonumber \\
 &&+\frac{1}{16(m_1+m_2+m_3)^5}\Bigl[
  4(m_1+m_2)\left(m_1^2{\cal L}_t(m_1,m_2,m_3)+m_2^2{\cal L}_t(m_2,m_1,m_3)
    \right)\nonumber \\
  &&{\kern+30pt}-m_1\log(m_1^2)\left(m_1^2+m_1m_2+m_2^2\right)
  -m_2\log(m_2^2)\left(m_1^2+m_1m_2+m_2^2\right)\nonumber \\
  &&{\kern+30pt}
 +\log(m_3^2)\left(m_1^3+2m_1^2m_2+2m_1m_2^2+m_2^3\right)\Bigr]\nonumber \\
 &&+\frac{1}{32(m_1+m_2+m_3)^4}\Bigl[
  -4m_1^2{\cal L}_t(m_1,m_2,m_3)-4m_2^2{\cal L}_t(m_2,m_1,m_3)\nonumber \\
  &&{\kern+120pt}+m_1\log(m_1^2)\left(3m_1+2m_2\right)
  +m_2\log(m_2^2)\left(2m_1+3m_2\right)\nonumber \\
  &&{\kern+120pt}-\log(m_3^2)\left(3m_1^2+4m_1m_2+3m_2^2\right)
  +2(m_1^2+m_1m_2+m_2^2)\Bigr]\nonumber \\
  &&{\kern-16pt}+\frac{1}{32(m_1+m_2+m_3)^3}\Bigl[
    -m_1\log(m_1^2)
  -m_2\log(m_2^2)
   +\left(m_1+m_2\right)\log(m_3^2)
 -2(m_1+m_2)\Bigr]\nonumber \\
 &&+ \frac{1}{64(m_1+m_2+m_3)^2}
 -\frac{\pi\sqrt{m_1m_2m_3(m_1+m_2+m_3)}}{4(m_1+m_2+m_3)^6}
 \left(m_1^2+m_1m_2+m_2^2\right)\nonumber \\
 &&+\frac{\pi\sqrt{m_1m_2m_3(m_1+m_2+m_3)}}{4(m_1+m_2+m_3)^5}
   \left(m_1+m_2\right)\nonumber \\
 &&+\frac{\sqrt{m_1m_2m_3(m_1+m_2+m_3)}}{64(m_1+m_2+m_3)^4}\Biggl\{
 \pi\Bigl[\frac{3}{2}\left(\log(m_1^2)+\log(m_2^2)+\log(m_2^2)\right)
 \nonumber \\
  &&-2(\log(m_1+m_2)+\log(m_1+m_3)+\log(m_2+m_3))+\log(m_1+m_2+m_3)-3
 \Bigr] \nonumber \\
 &&+ 2{{\cal I}_3(m_1,m_2,m_3)}  -2K\Biggr\}
 \ . \labbel{EX16} \end{eqnarray}

As we know only the expansion of \(\HC{(0,2)}\) around \(n=4\) and not
 its exact form for arbitrary \(n\), to proceed, we write also the
 expansion in \((n-4)\) of the complete first master amplitude \(\F{0}\).
 The singular and regular part are well distinct for continuous
 arbitrary \(n\); in the expansion around \(n=4\), the singular part gives
 terms proportional to \( \log(x)\) plus other terms , without  \( \log(x)\),
 which mix with the terms coming from the regular part.
 Expanding \Eq{EX1} in \((n-4)\) we find

 \begin{eqnarray}
 &&{\kern-15pt}\F{0} = C^2(n)\Biggl\{\nonumber \\
 &&\frac{1}{(n-4)^2}\GPI{(-2)}{0} + \frac{1}{(n-4)}\GPI{(-1)}{0}
 + \GPI{(0)}{0}\nonumber \\
 &&+ \ x \ \ \Biggl[
   \frac{1}{(n-4)}{\cal H}_{(0,1)}^{(-1)}
  + {\cal H}_{(0,1)}^{(0)}
   \Biggr]\nonumber \\
 &&+ x^2 \Biggl[{\cal H}_{(0,2)}^{(0)}
   -{\cal H}_{(0,2)}^{(-1)}\left(\log(x)
   +\frac{1}{2}\log\left(\frac{m_1m_2m_3}{(m_1+m_2+m_3)^3}\right)\right)
\nonumber \\
    &&{\kern+50pt}+ b \frac{\sqrt{m_1m_2m_3(m_1+m_2+m_3)}}{(m_1+m_2+m_3)^4}
    \Biggr]
   \ \ + \ \ O(n-4,x^3)\Biggr\}
 \ , \labbel{EX17} \end{eqnarray}

\noindent
where the constant \(b\) comes from the expansion of
 \( {\cal C}_{\cal H}\left(n\right)\) around \(n=4\)

\begin{eqnarray}
 {\cal C}_{\cal H}\left(n\right) = C^2(n)\left[
-\frac{1}{(n-4)} \ \frac{\pi}{32} + b \ + \ \ O(n-4) \right]
\ . \labbel{EX18} \end{eqnarray}

The requirement of the disappearance of the pole term in the coefficient of
 \(x^2\) in \Eq{EX17} and the knowledge of the pole term of \(\HC{(0,2)}\)
 fix the pole term in \( {\cal C}_{\cal H}\left(n\right) \).
 That requirement fix also the absence of higher pole terms in
 \( {\cal C}_{\cal H}\left(n\right) \).
 In the expansion \Eq{EX17} there are two still unknown constants:
 \(K\) in the expression of \({\cal H}_{(0,2)}^{(0)}\) and \(b\)
 from \(  {\cal C}_{\cal H}\left(n\right)\). They cannot be fixed
 separately due to the \(n=4\) expansion, but only in the occurring
 combination \( (b-K/32) \).
 As we are interested only in the \(n=4\) expansion of the master integrals,
 the knowledge of the combination of the constants appearing in \Eq{EX17}
 is enough to fix all the higher order terms in the threshold expansion.
 To fix the combination of the constants it is sufficient to know the term
\(\sim x^2\) of \( \F{0}\) for fixed values of the masses.
 Due to the factor in front of \(b\) and \(K\),
 \( \sqrt{m_1m_2m_3} \), all masses have to be different from zero
 and so the simplest choice is the equal mass case \( (m_1=m_2=m_3=m) \).
 As the analytical result for equal mass case is available \cite{DS},
 we limited ourselves to its numerical check, which can be performed
 by using the dispersion relation representation of the master integral
 \(F_{0}(n=4,m^2,m^2,m^2,p^2)\) \cite{CCR}.
 Its second derivative reads

 \begin{eqnarray}
  \frac{\partial^2 F_{0}(n=4,m^2,m^2,m^2,p^2)}{\partial (p^2)^2}
  =
\int\limits_{9m^2}^{\infty} du \frac{1}{8(u+p^2 )^3} \ E_0(u,m)
     \ ,
 \labbel{V5} \end{eqnarray}

\noindent
where
 \begin{eqnarray}
 E_0(u,m) =\frac{1}{u}
  \int\limits_{4m^2}^{(\sqrt{u}-m)^2}
  db \frac{ R(u,b,m^2)R(b,m^2,m^2)}{b} \ \ .
 \labbel{V5A} \end{eqnarray}

 At \(p^2=-9m^2\) the integral \Eq{V5} is logarithmically divergent
 (in agreement with \Eq{EX17}), but the divergent part can be
 easily extracted by integrating by parts, providing us with the
 two needed leading terms in the threshold expansion of the integral

  \begin{eqnarray}
   \frac{\partial^2 F_{0}(n=4,m^2,m^2,m^2,p^2)}{\partial (p^2)^2}
  &=&
   -\frac{\pi\sqrt{3}}{6^4m^2} \log\left(\frac{x_e}{m^2}\right)
 \nonumber \\
 &&{\kern-140pt}-\frac{1}{16}\int\limits_{9m^2}^{\infty} du
 \log\left(\frac{u-9m^2}{m^2}\right)\frac{\partial^3E_0(u,m)}{\partial u^3}
  + O(x_e)\labbel{V5B1}  \\
 & =& -\frac{\pi\sqrt{3}}{6^4m^2} \log\left(\frac{x_e}{m^2}\right)\nonumber \\
  &&{\kern-140pt}-\frac{1}{864m^2} +
 \frac{\pi\sqrt{3}}{m^2} \left( -\frac{1}{972} +\frac{\log(2)}{432}
    +\frac{\log(3)}{648} \right) -\frac{5\sqrt{3}}{1296m^2}
      \hbox{Cl}_2\left(\frac{\pi}{3}\right)+ O(x_e)
 \ ,
 \labbel{V5B}
 \end{eqnarray}

\noindent
 where \( x_e = p^2+9m^2\) corresponds to the equal mass limit
 \( m_1=m_2=m_3=m \) of \( x\) in \Eq{EX1A}
 and the explicit value of the integral is taken from \cite{DS}.
 The analytical result of \cite{DS},
 in \Eq{V5B}, agrees with the numerical value of the integral in \Eq{V5B1},
 so we did not reevaluate analytically the integral.

 For extracting the combination of the constants \( (b-K/32) \),
 we consider \Eq{EX17} in the equal mass limit \(m_1=m_2=m_3=m\),
 and we compare its second \( p^2\) derivative with \Eq{V5B}, obtaining

 \begin{eqnarray}
 b-\frac{K}{32} = \pi\left(-\frac{1}{32}
    +\frac{5}{32}\log(2)\right)
   +\frac{1}{8}\hbox{Cl}_2\left(\frac{\pi}{2}\right)
\ , \labbel{EX22} \end{eqnarray}

\noindent
 to be used in \Eq{EX17} and in all the other \(n=4\) expansions of the
 master integrals. For completeness we report \Eq{EX17} in the equal mass
 limit \(m_1=m_2=m_3=m\) in our notations
 \begin{eqnarray}
 &&F_0(n,m^2,m^2,m^2,p^2) = C^2(n)\Biggl\{
 -\frac{3m^2}{8(n-4)^2}
 + \frac{3m^2}{32(n-4)}\left(3-4\log(m^2) \right)
 \nonumber \\
 && {\kern+140pt}-\frac{ m^2\pi \sqrt{3}}{6}
    +\frac{3m^2}{128}\left(15+12\log(m^2)-8\log^2(m^2)\right)
 \nonumber \\
 &&+x_e \Biggl[ \frac{1}{32(n-4)}
  +\frac{\pi \sqrt{3}}{108} -\frac{23}{384} +\frac{1}{32}\log(m^2)
 \Biggr]\nonumber \\
 &&+ x_e^2 \frac{\sqrt{3}}{648m^2}\Biggl[  \pi \left(
 -\frac{1}{4}\log\left(\frac{x_e}{m^2}\right)
 +\frac{3}{4}\log(2)
 +\frac{1}{2}\log(3) +\frac{1}{24} \right)
 -\frac{5}{4}\hbox{Cl}_2\left(\frac{\pi}{3}\right)
  - \frac{\sqrt{3}}{8}
  \Biggr] \nonumber \\
 &&+ \ \ O(n-4,x_e^3) \Biggr\}
\ . \labbel{EX19} \end{eqnarray}

 The higher order terms in the
 expansion at threshold can be easily obtained algebraically, but they
 are of interest only in case one wants to calculate the
 master amplitudes using the threshold expansion \cite{DS}, which is not
 our purpose here. The expansion of all the master integrals
 up to terms \(\sim x\) is however necessary for the numerical solution
 of the system of equations obtained in \cite{CCLR}.

 The threshold expansion of the remaining master integrals can be obtained
 using \Eq{Fi}. One gets

\begin{eqnarray}
 \F{3} = &&C^2(n)\Biggl[
  \frac{1}{(n-4)^2}\GPI{(-2)}{3} + \frac{1}{(n-4)}\GPI{(-1)}{3}\nonumber \\
 &&+ \GPI{(0)}{3}
 + \ x \ \  {\cal H}_{(3,1)}^{(0)} \ \ + \ \ O(n-4,x^2) \Biggr]
\ , \labbel{EX20} \end{eqnarray}

\noindent
where \(\GPI{(i)}{j}\) are defined in \Eq{7} and \Eq{V22} and

\begin{eqnarray}
&&{\cal H}_{(3,1)}^{(0)} =
 -\pi \frac{\sqrt{m_1m_2m_3(m_1+m_2+m_3)}}{4m_3(m_1+m_2+m_3)^4}
  \left(m_1+m_2\right)
 \nonumber \\
&&+\frac{\sqrt{m_1m_2m_3(m_1+m_2+m_3)}}{8m_3(m_1+m_2+m_3)^3}
 \Biggl[
  \pi\Biggl( \frac{1}{2} \log\left(\frac{x}{(m_1+m_2+m_3)^2}\right)
  +1 +\frac{1}{2} \log\left(\frac{m_2+m_3}{m_1}\right)\nonumber \\
                    &&+\frac{1}{2} \log\left(\frac{m_1+m_3}{m_2}\right)
                    +\frac{1}{2} \log\left(\frac{m_1+m_2}{m_3}\right)
  \Biggr)
 -16 \left(b-\frac{K}{32}\right)
 -\frac{1}{2}{{\cal I}_3(m_1,m_2,m_3)}\Biggr]\nonumber \\
      &&+\frac{1}{8(m_1+m_2+m_3)^4}
 \left[m_1^2{\cal L}_t(m_1,m_2,m_3)+m_2^2{\cal L}_t(m_2,m_1,m_3)\right]
\nonumber \\
 &&+\frac{1}{16(m_1+m_2+m_3)^3}\left[
 m_1\log(m_1^2)+ m_2\log(m_2^2) -(m_1+m_2)\log(m_3^2)\right]\nonumber \\
 &&-\frac{1}{16(m_1+m_2+m_3)^2}
 \ . \labbel{EX21} \end{eqnarray}

The expansion of \(\F{1}\) and \(\F{2}\)  can be obtained by a permutation
of the masses.
%%%%%%%%%%%%%%%%%%%%%%%%%%%%%%%%%%%%%
 \section {Summary.}
 In this paper we have presented the expansion of the 2-loop sunrise selfmass
 master amplitudes at the threshold \(p^2 = -(m_1+m_2+m_3)^2\).
 We define the expansion in \Eq{EX1};
 the values of the amplitudes at the threshold are given
 in \Eq{6}, \Eq{7}, \Eq{V17} and \Eq{V22}.
 The first order terms in the threshold expansion of the master
 amplitudes at \(n=4\) are presented in  \Eq{EX5A}
 and \Eq{EX20}, while only for the first master amplitude
 \(F_{0}(n=4,m_1^2,m_2^2,m_3^2,p^2 = -(m_1+m_2+m_3)^2 )\)
 the  second order  term in the threshold expansion
 is presented in \Eq{EX17}.
 The higher order terms, which are not given explicitly here, can be easily
 found by solving recursively, at each order, a system of four algebraic
 linear equations. As said already in \cite{CCR},
 the expansion at the pseudothreshold cannot be simply
 deduced from the known expansion at the threshold  and vice versa,
even if at first sight they seem to be connected by the change of sign
\( m_3 \to -m_3 \). In fact the analytic properties of the amplitudes
are different at the two points: at the pseudothreshold
the sunrise amplitudes are regular, so that
the solution of the system of equations \cita{CCLR}
  can be expanded as a single power series,
while at the threshold the sunrise amplitudes develop a branch point
and its expansion is indeed the sum of two series \Eq{EX1}, \cita{I}.

\vskip 0.4 cm

%%%%%%%%%%%%%%%%%%%%%%%%%%%%%%%%%%%%%
{\bf Acknowledgments.}

One of us (HC) is grateful to the Bologna Section of INFN and to the Department
 of Physics of the Bologna University for support and kind hospitality.

%%%%%%%%%%%%%%%%%%%%%%%%%%%%%%%%%%%%%%%%%%%%%%%%%%%%%%%%%%%%%%%%%%%%%%%%
\def\NP{{\sl Nucl. Phys.}}
\def\PL{{\sl Phys. Lett.}}
\def\PR{{\sl Phys. Rev.}}
\def\PRL{{\sl Phys. Rev. Lett.}}
\def\NC{{\sl Nuovo Cim.}}

\end{document}